\documentclass[preprint,review,11pt]{elsarticle}
\usepackage[margin=1.5in]{geometry}

\usepackage{amssymb}

\journal{XXX}

\usepackage{graphicx}
\graphicspath{{./figures/}}  
\usepackage[table]{xcolor}
\usepackage{soul,xcolor}
\definecolor{light-gray}{gray}{0.9}
\usepackage{amsmath,amsfonts,amssymb,amsthm}
\usepackage[version=4]{mhchem}
\usepackage{siunitx}
\usepackage{longtable}
\usepackage{tabularx}
\usepackage{booktabs}
\usepackage{multirow}
\usepackage{cleveref}
\usepackage{subcaption}
\usepackage{enumitem}
\usepackage{todonotes}

\newcommand{\x}{{\textbf{\em x}}}
\newcommand{\xx}{{\textbf{\em y}}}
\newcommand{\zz}{{\boldsymbol\theta}}
\newcommand{\QQ}{{Q}}
\newcommand{\QQt}{{Q_{MFMC}}}

\newcommand{\SSSt}{{\widetilde S}}

\newcommand{\CC}{{C}}
\newcommand{\KKKt}{{\widetilde K}}
\newcommand{\KKK}{{K}}

\newcommand{\SSS}{{\mathcal S}}
\newcommand{\VVV}{{\it V}}
\newcommand{\VVVt}{{V}}

\newcommand{\ppp}{{B}}
\newcommand{\LLL}{{L}}
\newcommand{\ttt}{{t}}
\newcommand{\nnn}{{n}}

\renewcommand{\d}{\mathop{}\!\mathrm{d}}
\newcommand{\D}{\Omega}

\newcommand{\DI}{\D_{I}}
\newcommand{\DD}{\D\cup\DI}
\newcommand{\R}{\mathbb{R}}
\newcommand{\kernel}{\gamma}

\newcommand{\cker}{c_{\kernel}}

\newcommand{\cpot}{c_{F}}

\newcommand{\p}{u}

\definecolor{Max}{cmyk}{1.00, .08, 0, .28}
\definecolor{Olena}{cmyk}{0.67, 0, .57, .36}
\definecolor{Parisa}{cmyk}{.09, .29, .66, .24}

\begin{document}

\begin{frontmatter}



\title{Multifidelity Methods for Uncertainty Quantification of a Nonlocal Model for Phase Changes in Materials\footnote{Notice: This manuscript has been authored in part by UT-Battelle, LLC, under contract DE-AC05-00OR22725 with the US Department of Energy (DOE). The US government retains and the publisher, by accepting the article for publication, acknowledges that the US government retains a nonexclusive, paid-up, irrevocable, worldwide license to publish or reproduce the published form of this manuscript, or allow others to do so, for US government purposes. DOE will provide public access to these results of federally sponsored research in accordance with the DOE Public Access Plan (\texttt{http://energy.gov/downloads/doe-public-access-plan}).}}


\author[1]{Parisa Khodabakhshi\corref{cor1}}
\ead{pak322@lehigh.edu}
\author[2]{Olena Burkovska}
\author[3]{Karen Willcox}
\author[3]{Max Gunzburger}
\cortext[cor1]{Corresponding author}
\affiliation[1]{organization={Department of Mechanical Engineering \& Mechanics, Lehigh University},
city={Bethlehem},
state={PA 18015},
country={USA}}
\affiliation[2]{organization={Computer Science and Mathematics Division, Oak Ridge National Laboratory},
state={TN 37831},
country={USA}}
\affiliation[3]{organization={Oden Institute for Computational Engineering \& Sciences, The University of Texas at Austin},
state={TX 78712},
country={USA}}

\begin{abstract}
This study is devoted to the construction of a multifidelity Monte Carlo (MFMC) method for the uncertainty quantification of a nonlocal, non-mass-conserving Cahn-Hilliard model for phase transitions with an obstacle potential. We are interested in the estimation of the expected value of an output of interest (OoI) that depends on the solution of the nonlocal Cahn-Hilliard model. As opposed to its local counterpart, the nonlocal model captures sharp interfaces without the need for significant mesh refinement. However, the computational cost of the nonlocal Cahn-Hilliard model is higher than that of its local counterpart with similar mesh refinement, inhibiting its use for outer-loop applications such as uncertainty quantification. The MFMC method augments the desired high-fidelity, high-cost OoI with a set of lower-fidelity, lower-cost OoIs to alleviate the computational burden associated with nonlocality. Most of the computational budget is allocated to sampling the cheap surrogate models to achieve speedup, whereas the high-fidelity model is sparsely sampled to maintain accuracy. For the non-mass-conserving nonlocal Cahn-Hilliard model, the use of the MFMC method results in, for a given computational budget, about one-order-of-magnitude reduction in the mean-squared error of the expected value of the OoI relative to that of the Monte Carlo method.
\end{abstract}


\end{frontmatter}


\section{Introduction}

Phase-field models, governed by partial differential equations (PDEs), have been widely used to describe moving boundaries between phases without the need to explicitly track that boundary and instead replace the sharp boundaries with diffuse interfaces having a finite thickness. Typically, to approximate very thin or sharp interfaces with a diffuse-interface phase-field model, one needs to increase the mesh resolution in the vicinity of the interfaces to properly resolve their width. Increased mesh resolution results in higher computational costs associated with solving the corresponding system. 

In \cite{Burkovska_Gunzburger_2021}, a new nonlocal Cahn-Hilliard model that features a non-smooth obstacle potential is introduced that promotes a sharp interface {\em without the need for mesh refinement near the interface}. Later, these results were generalized for a non-isothermal case in~\cite{Burkovska_2023}. In contrast to local models where interactions occur through contact, nonlocal models feature a finite length scale $\delta$ (referred to as the {\em horizon}) where any two points separated by a distance smaller than the horizon interact with each other \cite{Gunzburger_Lehoucq_2010,Du_Gunzburger_Lehoucq_Zhou_2013,DElia_Du_Glusa_Gunzburger_Tian_Zhou_2020}. Even though nonlocal models allow for sharp solution discontinuities (e.g., sharp interfaces for phase-change models), the inherent distant interactions give rise to the reduced sparsity of discretized models and consequently incur higher computational costs compared to their local PDE counterparts with similar refinement. The effect of distant interactions on the increased computational expenses becomes more significant as the spatial dimension increases.

Multifidelity methods \cite{Peherstorfer_Willcox_Gunzburger_2016,Peherstorfer_Willcox_Gunzburger_2018} reduce the computational cost of many-query applications (e.g., uncertainty quantification, optimization, inverse problems) by augmenting a high-fidelity model of interest with a set of lower-fidelity surrogate models that incur lower computational expenses. In the MFMC method, for a given computational budget, most of that budget is allocated to the low-fidelity models to achieve speedup, and the high-fidelity model is sampled to avoid biases and preserve accuracy. 

In \cite{Khodabakhshi_Willcox_Gunzburger_2021}, the multifidelity Monte Carlo (MFMC) method described in \cite{Peherstorfer_Willcox_Gunzburger_2016} is used for uncertainty quantification in a nonlocal diffusion model setting. In that setting, the surrogate models are built by coarsening the grid, reducing the horizon $\delta$, or both. The MFMC method alleviates the burden resulting from the increases in the computational costs incurred by nonlocal models relative to those for related local PDE models. In this paper, we are interested in the uncertainty quantification of an output of interest (OoI) that requires multiple solutions of the generalized nonlocal Cahn-Hilliard model with the obstacle potential considered in \cite{Burkovska_Gunzburger_2021,Burkovska_2023}. As such, we apply the MFMC approach of \cite{Peherstorfer_Willcox_Gunzburger_2016} to the nonlocal non-mass-conserving Cahn-Hilliard model to reduce the computational costs while maintaining the desired accuracy.

The structure of the paper is as follows. In Section \ref{section2}, we provide a brief description of the generalized nonlocal Cahn-Hilliard model. In Section \ref{section3}, we review the MFMC method given in \cite{Peherstorfer_Willcox_Gunzburger_2016,Peherstorfer_Willcox_Gunzburger_2018}. In Section \ref{section4}, computational illustrations are provided that result from the application of the MFMC method described in Section \ref{section3} to the Cahn-Hilliard model described in Section \ref{section2}.
Concluding remarks are given in Section \ref{section5}.


\section{Generalized Nonlocal Cahn-Hilliard Model}\label{section2}

The local Cahn-Hilliard model introduced in~\cite{Cahn_Hilliard_1958} describes the phase separation of a binary mixture, e.g., a binary alloy. An auxiliary variable $\p(x)$ varying within the range $\left[-1,\,1\right]$, referred to as the order parameter, is introduced to describe the time evolution of the phase change, where the values $\p= 1$ and $-1$ represent the two pure phases, e.g., solid or liquid in the solidification problem.

The local Cahn-Hilliard model can be derived as a gradient flow of the Ginzburg-Landau energy
\[
E(u)=\int_{\D} \frac{\varepsilon^2}{2}|\nabla u|^2\d x +\int_{\D} F(u)\d x,
\]
which results in \eqref{eq:CH_local} used to describe the time-evolution of the order parameter $u$ 
\begin{equation}\label{eq:CH_local}
\begin{cases}
\partial_t u-\Delta w=0\\
w=-\varepsilon^2 \Delta u+F^\prime(u).
\end{cases}
\end{equation}
Here, $w$ is the chemical potential, the parameter $\varepsilon>0$, called an interface parameter, is related to the width of the diffuse interface, $F(u)$ is a double-well potential that assumes its minima at $\pm 1$, and $F'(u)$ is the derivative of $F(u)$ with respect to $u$. 
A common choice for $F$ is a smooth double-well potential given by 
\[
F(\p)=\frac{1}{4}(\p^2-1)^2.
\]
Alternative choices include logarithmic and obstacle potentials.
Being simple in implementation, smooth potentials, as defined above, allow for infeasible $u$ (i.e., for $u$ having values with a magnitude greater than unity). Even though logarithmic potentials produce feasible $u$, they exclude pure phases, that is $u \in (-1,1)$, $u\neq \pm 1$. In the nonlocal model considered in this work, we adopt an obstacle potential, which not only produces feasible solutions, $\p\in [-1,1]$ including pure phases but also allows for sharp interfaces during the temporal evolution of the phases in the nonlocal model. See \cite{Burkovska_Gunzburger_2021,Burkovska_2023} for more detailed comparisons of the use of smooth, logarithmic, and obstacle potentials. 

More specifically, an obstacle potential considered in this work is defined as follows
\begin{equation}\label{eq:potential_obstacle}
F_{\rm obs}(\p)=\left.
\begin{cases}
 {\displaystyle  \frac{\cpot}{2}}(1-\p^2),\quad &|\p|\leq 1\\
   +\infty,\quad &|\p|>1
\end{cases} \right\}
=\frac{\cpot}{2}(1-\p^2)+\mathbb{I}_{[-1,1]}(\p),
\end{equation}
where $\cpot>0$ and $\mathbb{I}_{[-1,\,1]}$ denotes the convex indicator function that ensures that the order parameter $u$ remains within the admissible range $\left[-1,\,1\right]$. While this potential enforces the solution to be always admissible, due to the presence of the indicator function, \eqref{eq:potential_obstacle} is non-smooth and is not differentiable in a classical sense. Thus, the derivative of $F(\p)$ in~\eqref{eq:CH_local} should be replaced by a generalized subdifferential 
\begin{align*}
    \partial F_{\rm obs}=-\cpot\p+\partial \mathbb{I}_{[-1,1]}(\p) =-\cpot\p+
    \begin{cases}
        (-\infty,0]\quad &\text{if}\;\p=-1\\
        0\quad &\text{if}\;\p\in (-1,1)\\
        [0,+\infty)\quad &\text{if}\;\p=1
    \end{cases}
\end{align*}
where $\partial \mathbb{I}_{[-1,1]}(\p)$ denotes the subdifferential of the indicator function $\mathbb{I}_{[-1,1]}(\p)$.

\subsection{Nonlocal model}
In this work, we study a nonlocal variant of the Cahn-Hilliard model~\eqref{eq:CH_local}, given by a generalized nonlocal Cahn-Hilliard model (which is an isothermal case of the  model in~\cite{Burkovska_2023}, and a variant of the Cahn-Hilliard model studied previously in \cite{Burkovska_Gunzburger_2021}) with an obstacle potential given by 
\begin{equation}\label{eq:nonl_CH_nm}
\begin{cases}
\partial_t u+\beta_1 w-\beta_2\Delta w=0\\
w=\LLL u+\partial F_{\rm obs}(u),
\end{cases}
\end{equation}
where $\LLL u$ denotes the nonlocal operator and $\beta_1,\beta_2\geq 0$ are parameters that control the nature of the temporal evolution of the phases.
For $\beta_1=0$ and $\beta_2>0$, the above model is a nonlocal mass-conserving Cahn-Hilliard model, which is an analog of the local system~\eqref{eq:CH_local}, whereas for $\beta_1>0$ and $\beta_2>0$ the resulting system is a non-mass conserving nonlocal Cahn-Hilliard model.
For $\beta_1=1$, $\beta_2=0$, \eqref{eq:nonl_CH_nm} results in a nonlocal variant of the Allen-Cahn model 
in \cite{Allen_Cahn_1979} which does not conserve mass and represents phase change in materials.
We limit our attention to the cases with $\beta_1> 0$ and $\beta_2>0$. These cases result in the nonlocal non-mass-conserving Cahn-Hilliard model that can produce solutions with sharp interfaces which evolve with time~\cite{Burkovska_Gunzburger_2021,Burkovska_2023}.

Let $\Omega\subset\R^\nnn$, $\nnn=1,2,3$, denote an $n$-dimensional physical domain. The nonlocal operator $\LLL$ is defined as follows
\begin{equation}\label{eq:nonl_operator}
\LLL\p:=\int_{\DD}(\p(\x)-\p(\xx))\kernel(\x-\xx)\d\xx=\cker \p(\x)-(\kernel*\p)(\x),\quad\x\in\D,
\end{equation}
where  $*$ represents convolution integral, $(\kernel*u)(\x)=\int_{\DD}\p(\xx)\kernel(\x-\xx)\d\xx$, $\cker(\x)=(\kernel*1)(\x)=\int_{\DD}\kernel(\x-\xx)\d\xx$, and $\cker(\x)=\cker$ is constant for all $\x\in\D$.

In \eqref{eq:nonl_operator}, $\kernel\colon\R^n\to\R^{+}$ denotes a kernel function that serves the dual roles of incorporating a constitutive function and accounting for the distant interactions. The kernel is assumed to be an integrable and compactly supported radial function where $\kernel(\xx-\x)={\kernel}(|\xx-\x|)$. The support of the kernel is dictated by the horizon parameter $\delta>0$ (i.e., $\kernel(|\xx-\x|)=0$ if $|\xx-\x|>\delta$). Within its support, $\kernel(|\xx-\x|)$ is positive except that we allow for $\kernel(|\xx-\x|)= 0$ for $|\xx-\x|=\delta$.
In addition, $\Omega_I$ in \eqref{eq:nonl_operator} denotes the interaction region consisting of points outside of the physical domain $\Omega$ which are less than $\delta$ apart from any points in $\Omega$, i.e., 
$$
\Omega_I:=\{\xx\in\R^n\setminus\Omega \,\colon\,\, |\xx-\x|\le\delta \,\,\, \mbox{for some}\,\,\, \x\in\Omega\}.
$$

A local model, such as~\eqref{eq:CH_local}, can be considered as an approximation of the nonlocal model when the horizon parameter vanishes, $\delta\to 0$. In this case, under appropriate scaling of the kernel, it can be shown that $L$ converges to a second-order elliptic operator, e.g., for a specific constant media property $L\p\to -\varepsilon^2\Delta\p$ when $\delta\to 0$; see, e.g.,~\cite{Du_Ju_Li_Qiao_2018}.

In~\cite{Burkovska_Gunzburger_2021}, it is shown that under suitable conditions on the kernel and other model parameters, the nonlocal mass-conserving Cahn-Hilliard model~\eqref{eq:nonl_CH_nm} with $\beta_1= 0$, $\beta_2>0$ can produce sharp-interfaces in the solution having only pure phases if the parameter
\begin{equation}\label{eq:xi_constant}
 \xi:=\cker-\cpot=\int_{\Omega\cup\Omega_I}\kernel(\x-\xx)\d\xx-c_F
\end{equation}
vanishes. Sharp interfaces are also preserved (see \cite{Burkovska_2023}) for the non-mass-conserving case ($\beta_1>0$, $\beta_2>0$) provided that $\xi=0$.
For positive values, $\xi$ plays the role of a nonlocal interface parameter whose value dictates the width of the interface, similar to the local interface parameter $\varepsilon>0$. In contrast to the local case, we can set $\xi$ to zero, which results in sharp interfaces with non-vanishing nonlocal interactions, whereas the sharp interfaces in the local model are only achieved in the limiting case $\varepsilon\to 0$. To promote sharp (or nearly sharp) interfaces in the solution of the nonlocal Cahn-Hilliard model, we choose the model parameters such that $\xi$ vanishes or becomes sufficiently small. 

\begin{figure}[h!]
\begin{center}
\begin{tabular}{ccc}
\includegraphics[width=.3\textwidth]{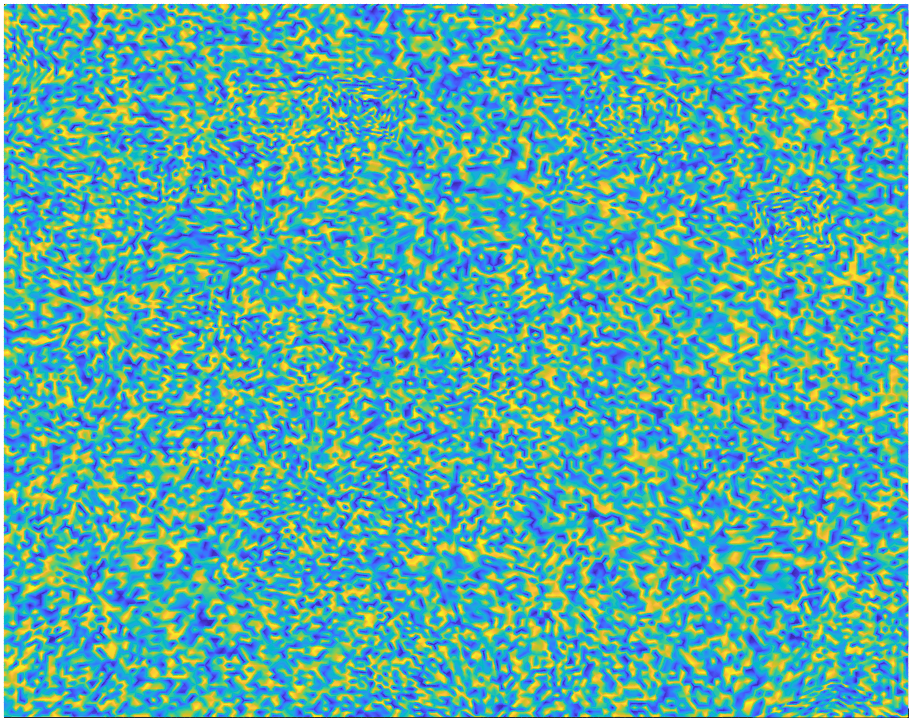}
&
\includegraphics[width=.3\textwidth]{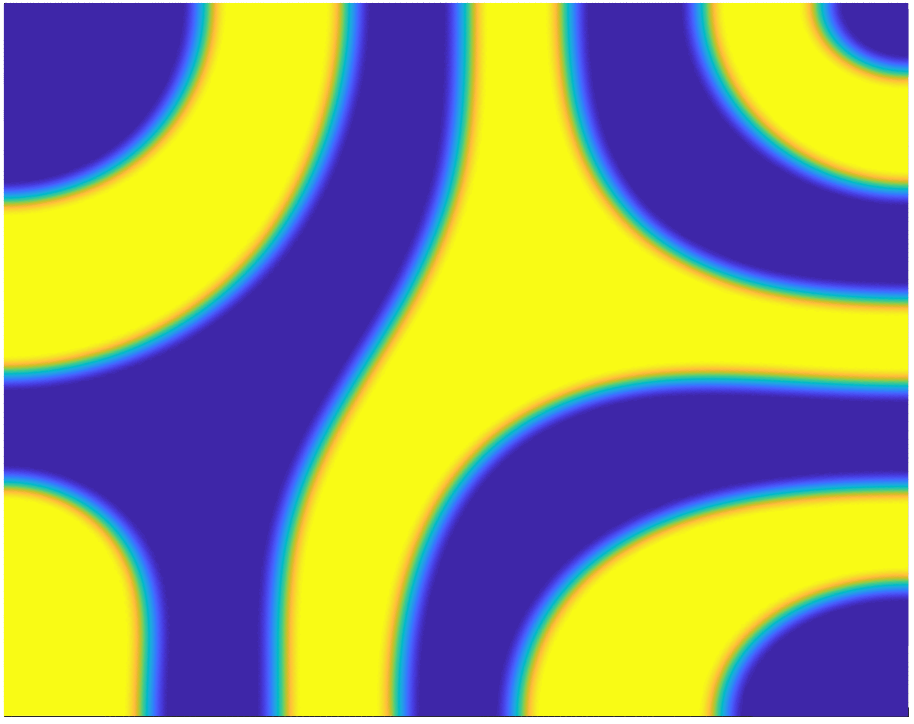}
&
\includegraphics[width=.3\textwidth]{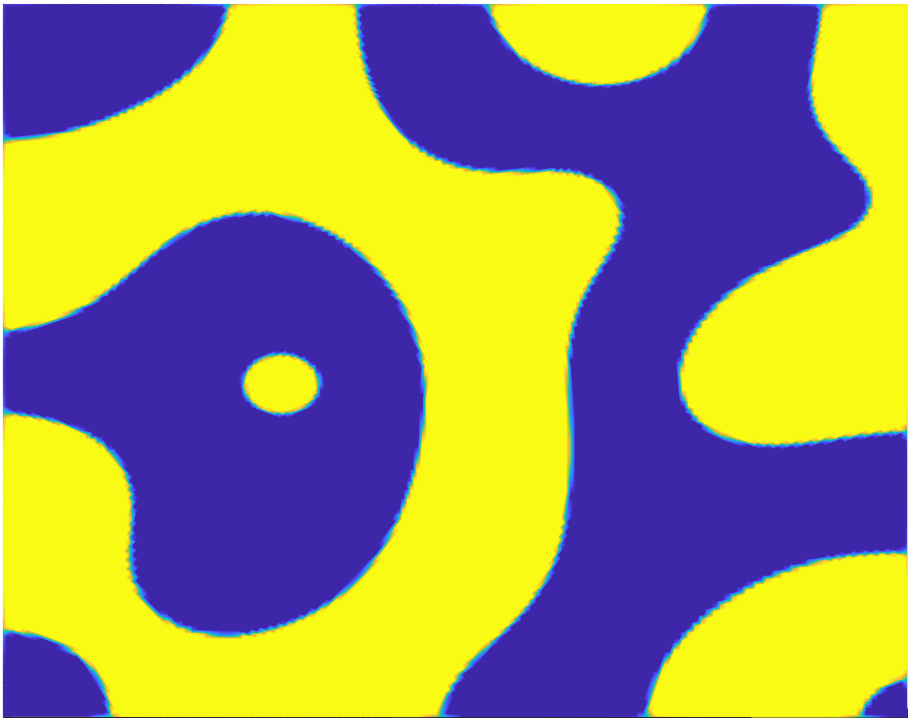}
\\
\small{random initial condition}
&
\small{local solution}
&
\small{nonlocal solution}
\\[-4ex]
\end{tabular}
\end{center}
\caption{
A random initial condition sampled at the grid nodes evolves at a later time to a local solution having diffuse interfaces whereas, for the same spatial grid and at the same later time, evolves into a nonlocal solution having sharp interfaces.}
\label{fig:figure1}
\end{figure}

For the same initial condition, the same spatial grid, the same finite element discretization of \eqref{eq:nonl_CH_nm} and of its related local PDE model, and the same time steps, Figure \ref{fig:figure1} illustrates the difference between the evolution of the interfaces for the local and nonlocal mass-conserving Cahn-Hilliard equations. In particular, the fuzzy interface for the local model extends over several grid cells whereas, even though the same grid and same time steps are used for the nonlocal model, the interface straddles across only one or at most two grid cells. 

\section{A Multifidelity Monte Carlo Method}\label{section3}

In this section, we provide a brief exposition of the MFMC method introduced in \cite{Peherstorfer_Willcox_Gunzburger_2016,Peherstorfer_Willcox_Gunzburger_2018} which is based on augmenting the high-fidelity model with a set of lower-fidelity models to determine a quantity of interest (QoI). In Section \ref{section4}, the MFMC method will be applied to the setting of the generalized nonlocal Cahn-Hilliard model of \eqref{eq:nonl_CH_nm} to address and alleviate the sparsity-related increased computational costs compared to its local counterpart. The MFMC method has been shown to be effective in other settings for significantly reducing the cost while preserving the accuracy of determining an estimator.

In our problem setup, we have in hand a model of interest, which we refer to as the {\em high-fidelity model}, that maps a vector of random input parameters $\zz$ to the OoI $f_1(\zz)=f_1\big(u(\x,T;\zz)\big)$. The OoI depends on the solution $u(\x,T;\zz)$ of \eqref{eq:nonl_CH_nm} at a chosen later time $T>0$. The high-fidelity model is used to define a QoI corresponding to the OoI $f_1(\zz)$. Specifically, our goal is to estimate the expectation of the OoI $f_1(\zz)$ using random samples of the parameter vector $\zz$
\begin{equation}\label{eq:expectation}
\text{QoI}:\;\QQ = {\mathbb E}[f_1(\zz)].
\end{equation}
In practice, a common means for effecting an approximation of \eqref{eq:expectation} is Monte Carlo estimation, i.e., we have the Monte Carlo estimator $\QQ_{MC}$ given by
\begin{equation}\label{eq:mc}
\QQ_{MC} = \frac1{M_{MC}} \sum_{m=1}^{M_{MC}} f_1(\zz_m)
\approx
\QQ = \mathbb{E}\left[f_1(\mathbf{\zz})\right] .
\end{equation}
Should the evaluation of $f_1(\zz)$ be an expensive endeavor, the slow convergence of the Monte Carlo estimator as the number of samples $M_{MC}$ of $\zz$ increases, would result in high, perhaps prohibitive, computational costs.

In Section \ref{modelselection}, we elaborate on how to optimally select particular surrogate models from a set of surrogate models to augment the high-fidelity model to be used in the multifidelity setting, and, in Section \ref{estimator32}, we consider the construction of the unbiased MFMC estimator for the QoI of \eqref{eq:expectation}.

\subsection{Model selection strategy}\label{modelselection}

In this section, we determine the optimal choices of surrogate models from a pool of available low-fidelity models to use to define an MFMC estimator. Assume that we have in hand a set $\SSSt = \big\{f_1(\zz),f_2(\zz),\ldots,f_\KKKt(\zz)\big\}$ consisting of the high-fidelity model $f_1(\zz)$ and also $\KKKt-1$ lower-fidelity surrogate models $f_k(\zz)$, $k=2,\ldots,\KKKt$, for which approximations of $u(\x,\ttt;\zz)$ are presumably less costly to obtain compared to that for the high-fidelity model. For the members of the set $\SSSt$, we determine
\begin{itemize}[topsep=0pt,parsep=-1pt,leftmargin=*]
    \item[--] $\sigma_k^2={\text{Var}}[f_k(\zz)]$ $\,\,\Longleftarrow\,\,$ the variance of model $f_k$ for $k=1,\ldots,\KKKt$,
    \item[--] $\rho_{1,k} = {\text{Cov}}[f_1(\zz),f_k(\zz)]/\sigma_1\sigma_k$ $\,\,\Longleftarrow\,\,$ the Pearson correlation coefficient between the models $f_1(\zz)$ and $f_k(\zz)$ for $k=1,\ldots,\KKKt$,
    \item[--] $\CC_k$ $\,\,\Longleftarrow\,\,$ the cost of model $f_k$ for $k=1,\ldots,\KKKt$.
\end{itemize}
\vspace{3pt}

\noindent
Without loss of generality, the models can be ordered to have Pearson correlation coefficients in decreasing order, i.e., we have that
\begin{equation}\label{eq:conditiona}
1=\rho_{1,1}^2> \rho_{1,2}^2 > \rho_{1,3}^2 > \cdots > \rho_{1,\KKKt}^2,
\end{equation}
because if at first they are not so ordered, they can obviously be rearranged to satisfy \eqref{eq:conditiona}. The set $\SSSt$ contains $\KKKt^{subsets}= 2^{\KKKt-1}-$1 {\em distinct subsets} $\big\{\SSSt_1,\SSSt_2,\ldots,\SSSt_{\KKKt^{subsets}}\big\}=\SSSt^{subsets}$ where for $k\in\{1,\ldots,\KKKt^{subsets}\}$, the members of each subset $\SSSt_k\in\SSSt^{subsets}$ consist of the high-fidelity model $f_1(\zz)$ and anywhere from one to $\KKKt-1$ surrogate models. The membership of the models included in any subset $\SSSt_k$ is distinct from that of any other subset, and  the members inherit the ordering of the members of $\SSSt$ (i.e., they are ordered according to decreasing correlation coefficients). 

We next remove, for $k\in\{1,\ldots,\KKKt^{subsets}\}$, all subsets $\SSSt_k$ from $\SSSt$ having an index $j\in\{2,\cdots,|\SSSt_k|\}$ that violate the condition
\begin{equation}\label{eq:conditionb}
  \dfrac{\CC_{i_{j-1}}}{\CC_{i_{j}}}> \dfrac{\rho_{1,i_{j-1}}^2-\rho_{1,i_j}^2}{\rho_{1,i_j}^2-\rho_{1,i_{j+1}}^2}, \qquad j\in\{2,\cdots,|\SSSt_k|\},
\end{equation}
where $|\SSSt_k|$ is the cardinal number of the subset $\SSSt_k$, $i_j$ is the index of the $j$th model of subset $\SSSt_k$ when ordered according to \eqref{eq:conditiona}, and  $\rho_{1,i_{j+1}}=0$ for $j=|\SSSt_k|$. 
For example, if $\SSSt_k=\{1,3,9\}$, \eqref{eq:conditionb} should be checked for $j=2$ and 3 where for $j=2$, the indices $i_{j-1}$, $i_j$ and $i_{j+1}$ are 1, 3 and 9, respectively and for $j=3$ we have $i_{j-1}=3$, $i_j=9$ and $\rho_{1,i_{j+1}}=0$. Note that since the first member of each subset in $\SSS^{subsets}$ is the high-fidelity model $f_1$, the equality $C_{i_1}=C_1$ holds for all subsets $|\SSSt_k|$.

Condition \eqref{eq:conditionb} ensures that any subset $\SSSt_k\subseteq\SSSt$ including surrogate models which are too costly and/or are too poorly correlated with the high-fidelity model is not considered in the MFMC method, i.e., failure to satisfy condition \eqref{eq:conditionb} for any of the surrogate models within a subset will disqualify that subset from usage in the MFMC estimator.

We then collect all the remaining subsets, each of which has satisfied both \eqref{eq:conditiona} and \eqref{eq:conditionb}, into the set $\SSS^{subsets}=\{\SSS_1,\ldots,\SSS_{K^{subsets}}\}$ where $K^{subsets}$ is the number of remaining subsets.
Here, we have re-indexed the members of the set $\SSS^{subsets}$, i.e., we have that $\SSS_k\in\SSS^{subsets}$ with index $k$ is the same as $\SSSt_k\in\SSSt^{subsets}$ but with a different index $k$. 

We illustrate the above discussion by considering a case for which $\KKKt=4$. We are given a set $\SSSt = \{f_1, f_2,f_3, f_4\}$ that consists of the high-fidelity model $f_1$ and three surrogate models, which are ordered according to \eqref{eq:conditiona}. We then have that $\KKKt^{subsets}=7$ and  $\SSSt^{subsets}=\{ \SSSt_1,\ldots,\SSSt_7\}$.
\begin{center}
\begin{tabular}{ccccccc}
$\SSSt_1$ & $\SSSt_2$ & $\SSSt_3$ & $\SSSt_4$ & $\SSSt_5$ & $\SSSt_6$ & $\SSSt_7$ 
\\
$\overbrace{\{f_1,f_2\}}_{}$
&$\overbrace{\{f_1,f_3\}}_{}$
&$\overbrace{\{\mbox{\bf f}_1,{\mathbf f}_4\}}_{}$
&$\overbrace{\{f_1,f_2,f_3\}}_{}$
&$\overbrace{\{{\mathbf f}_1,{\mathbf f}_2,{\mathbf f}_4\}}_{}$
&$\overbrace{\{f_1,f_3,f_4\}}_{}$
&$\overbrace{\{f_1,f_2,f_3,f_4\}}_{}$
\\[-5ex]
\phantom{xx}
\end{tabular}
\end{center}

\noindent Suppose that the two subsets highlighted in boldface fail to satisfy \eqref{eq:conditionb}. Removing those two subsets ($K^{subsets}=5$), we are left with the the set $\SSS^{subsets}=\{ \SSS_1,\ldots,\SSS_5\}$ with
\begin{center}
\begin{tabular}{ccccc}
$\SSS_1$ & $\SSS_2$ & $\SSS_3$ & $\SSS_4$ & $\SSS_5$  
\\
$\overbrace{\{f_1,f_2\}}_{}$
&$\overbrace{\{f_1,f_3\}}_{}$
&$\overbrace{\{f_1,f_2,f_3\}}_{}$
&$\overbrace{\{f_1,f_3,f_4\}}_{}$
&$\overbrace{\{f_1,f_2,f_3,f_4\}}_{}$
\\[-5ex]
\phantom{xx}
\end{tabular}
\end{center}
\noindent where the sets have be re-indexed, e.g., $\SSS_3=\SSSt_4$.

Finally, for $k\in\{1,\ldots,\KKK^{subsets}\}$, we determine the variance reduction ratio $\VVV_k$ defined as
\begin{equation}\label{eq:variance_reduction_ratio}
    \VVV_k = \left(\sum_{j=1}^{|\SSS_k|}\sqrt{\frac{C_{i_j}}{C_1}(\rho_{1,i_j}^2-\rho_{1,i_{j+1}}^2)}\right)^2, 
\end{equation}
where the sum ranges over the indices $j$ corresponding to the members in the set $\SSS_k$. When selecting the optimal subset from $\SSS^{subsets}=\{\SSS_1,\ldots,\SSS_{K^{subsets}}\}$, the natural choice would be to choose the subset $\SSS_k$ having the smallest variance reduction ratio $\VVV_k$ as determined by \eqref{eq:variance_reduction_ratio}. However, other contributing factors will be discussed in Section \ref{estimator32}.

\subsection{The multifidelity Monte Carlo estimator}\label{estimator32}

For a subset $\SSS\subset\SSS^{subsets}$, the MFMC estimator $\QQt$ of $\QQ$ has the form of the telescopic sum
\begin{equation}\label{eq:MFMC_estimator}
  \QQt = \QQ_{m_1}^{(i_1)} + \sum_{j=2}^{|\SSS|} {\alpha_j\left( \QQ_{m_j}^{(i_j)} - \QQ_{m_{j-1}}^{(i_j)}\right)} \approx
  \QQ =  \mathbb{E}\left[f_1(\mathbf{z})\right],
\end{equation}
where $\QQ_{m_{\ell}}^{(i_j)}$ denotes the Monte Carlo estimator of the $j$th model of subset $\SSS$ using $m_\ell$ samples (i.e., $i_j$ is the index for the $j$th model in subset $\SSS$ when models are ordered in the order of descending correlation coefficient).

The number of samples $m_j$, $j=1,\ldots,|\SSS|$ and the coefficients $\alpha_j$, $j=2,\ldots,|\SSS|$ are determined via an optimization problem with a closed-form solution \cite{Peherstorfer_Willcox_Gunzburger_2016}. Consider the subset $\SSS=\{f_{i_1}(\zz),\ldots,f_{i_{|\SSS|}}(\zz)\}\subset\SSS^{subsets}$ constructed in Section \ref{modelselection} consisting of a high-fidelity model $f_{i_1}(\zz)=f_1(\zz)$ and $|\SSS|-1$ surrogates $f_{i_j}(\zz)$ for $j=2,\ldots,|\SSS|$. The coefficients $\alpha_j$, over-sampling ratios $r_j$, and the number of samples $m_j$ are determined as follows:
\begin{align}
    & \alpha_{j} = \dfrac{\rho_{1,i_j}\sigma_1}{\sigma_{i_j}} \quad\text{for}\;\; j=2,\ldots,|\SSS| \label{eq:alpha}\\
    & r_j = \left(\dfrac{C_1(\rho_{1,i_j}^2-\rho_{1,i_{j+1}}^2)}{ C_{i_j}(1-\rho_{1,i_2}^2)}\right)^{1/2} \;\;\text{for}\;\; j=1,\ldots,|\SSS| \label{eq:rk}\\
    & m_1 =\dfrac{\ppp}{\sum_{j=1}^{|\SSS|} C_{i_j}r_j}\quad\text{and}\quad m_j = m_1r_j\;\;\text{for} \;\; j=1,\ldots,|\SSS|. \label{eq:samples}
\end{align}
where $B$ denotes the computational budget.

To guarantee an unbiased estimator, the high-fidelity model must be sampled at least once. It can be shown that according to \eqref{eq:samples}, the minimum possible budget $\ppp_{{min}}$ for subset $\SSS\subset\SSS^{subsets}$ that guarantees $m_1\ge1$ must satisfy
\begin{equation}\label{eq:bmin}
\ppp_{{\min}}\geq \sum_{j=1}^{|\SSS|}{C_{i_j}r_j}
    = \frac{C_1}{\sqrt{1-\rho^2_{1,i_2}}} \sum_{j=1}^{|\SSS|}{\sqrt{\frac{C_{i_j}}{C_1}(\rho^2_{1,i_j}-\rho^2_{1,i_{j+1}})}} = C_1\sqrt{\frac{\VVV}{1-\rho^2_{1,i_2}}}.
\end{equation}
where $\VVV$ is the variance reduction ratio for subset $\SSS$.

\vskip5pt
\noindent
{\bf Remark.} The closed-form solution \eqref{eq:samples} of the optimization problem results in non-integer number of samples $m_j$, $j=1,\ldots, |\SSS|$. To ensure that the cost of the MFMC estimator is less than or equal to the computational budget $\ppp$, the number of samples is rounded down to the nearest integer number $\lfloor m_j \rfloor$. We henceforth treat the sample number $m_j$ as integers.
\quad$\Box$
\vskip5pt

As mentioned in Section \ref{modelselection}, the subset $\SSS\in\SSS^{subset}$ having the smallest variance reduction ratio $\VVV$ determined by \eqref{eq:variance_reduction_ratio} is considered the optimal choice for the MFMC estimation. However, to avoid any bias, the MFMC method requires that the high-fidelity model $f_1(\zz)$ be sampled at least once, i.e., $m_1\geq 1$.
The largest Pearson correlation coefficient $\rho_{1,i_2}$ between the high-fidelity model and the first surrogate model among the ordered surrogate models also strongly influences the minimum required computational budget for an unbiased MFMC estimation. The example subsets treated in Section \ref{section4} provide further insights concerning the selection of subsets for MFMC estimation.

We now turn to how the {\em validation} of the MFMC estimator can be effected through the use of the high-fidelity model $f_1(\zz)$. The validation tool we use is the mean-squared error (MSE) incurred by the MFMC estimator $Q_{MFMC}$ relative to the exact QoI $Q={\mathbb E}[f_1(\zz)]$. In general, one does not know that exact value so that $Q$ itself has to be estimated. That {\em off-line} task is effected by averaging the Monte Carlo estimator $Q_{MC}$ over many parameter samples $\zz$ to obtain an approximation $Q_{validation}$ of $Q$ such that the difference $Q_{validation}-Q$ is dominated by the error in $Q-Q_{MFMC}$. The MSE is then approximated by averaging the approximate squared error over a few, e.g. 10, different $Q_{MFMC}^\ell$ estimators, i.e., we have that
\begin{equation}\label{eq:mse_error}
   \mbox{MSE} \approx  \frac{1}{10}\sum_{\ell=1}^{10}{\left(Q_{validation}-Q_{MFMC}^\ell\right)^2} .
\end{equation}
If one is not satisfied with the accuracy obtained for the MFMC estimator, one has to raise the computational budget $B$. The theoretical MSE of the MFMC estimator is determined in \cite{Peherstorfer_Willcox_Gunzburger_2016} to be
\begin{equation}\label{eq:error_theory}
    \mbox{MSE}_{theoretical}
    = \frac{\sigma_1^2(1-\rho^2_{1,i_2})}{(m_1^*)^2C_1}B,
\end{equation} 
where $m_1^*$ denotes the raw value for the number of high-fidelity samples for budget $B$ before rounding down to the nearest integer.

In Section \ref{section4}, we compare the MSEs incurred by several MFMC estimators with that for the MC estimator. For MC estimation, we have that $\mbox{MSE}_{MC} \approx  \frac{1}{10}\sum_{\ell=1}^{10}{\left(Q_{validation}-Q_{MC}^\ell\right)^2}$.

\section{Numerical Results}\label{section4}
In this paper, we are interested in the uncertainty quantification of an OoI for the generalized nonlocal Cahn-Hilliard model~\eqref{eq:nonl_CH_nm} using the MFMC method. As the OoI, we consider the mass fraction of the phase $u=-1$ at a specified time $T>0$, which is given by
\begin{equation}\label{eq:ooi}
   f_1=\frac{1}{2}\left(1+\frac{1}{|\Omega|}\int_{\Omega}{u(\x,T)d\x}\right), 
\end{equation}
where $|\Omega|$ denotes the volume of $\Omega$ and $f=0$ and 1 correspond to 100\% of phases $\p=-1$ and $\p=1$, respectively, over the entire physical domain. The QoI we consider is the expected value of the OoI for given random initial conditions. The OoI \eqref{eq:ooi} is representative of, for example, the extent of tumor growth or of the amount of solidified material in the solidification of metals.

We set the computational domain to be a unit square in $\mathbb{R}^2$, $\D=(0,1)\times (0,1)$.
For the spatial and temporal discretization of the model~\eqref{eq:nonl_CH_nm} we follow the discretization approach studied in~\cite{Burkovska_Gunzburger_2021,Burkovska_2023}. We employ a piece-wise linear continuous finite element method using a mass lumping approach for spatial discretization. For temporal discretization, we adopt an implicit Euler time-stepping scheme with an explicit discretization of the convolution term $\kernel*\p^{k-1}$. 

In the numerical experiments, we choose the Gaussian radial kernel given by
\begin{equation}\label{eq:kernel_gauss}
\kernel(\x,\xx)=
\frac{4\varepsilon^2}{\pi(\delta_{hf}/3)^3}e^{-{|\x-\xx|^2}/{(\delta_{hf}/3)^2}}
{\mathcal X}_{B_\delta(\x)}(\xx)
=
\begin{cases}
\frac{4\varepsilon^2}{\pi(\delta_{hf}/3)^3}e^{-{|\x-\xx|^2}/{(\delta_{hf}/3)^2}}\quad&\text{if}\;\;|\x-\xx| \leq\delta\\
0\quad&\text{otherwise},
\end{cases}
\end{equation}
where ${\mathcal X}_{B_\delta(\x)}(\xx)$ denotes the indicator function for the Euclidean ball centered at $\x$ and having horizon $\delta$.
The scaling constant in the above kernel includes a parameter $\varepsilon^2$ in order to match the local model in the limit of vanishing nonlocal interactions, i.e., $L\p\to -\varepsilon^2\Delta\p$ for $\delta\to 0$. 
We set the value of the local interface parameter $\varepsilon$ such that the nonlocal interface parameter $\xi$ defined in \eqref{eq:xi_constant} is close to zero for the given $\delta_{hf}$ and we can achieve sharp interfaces in the numerical solution. More specifically, we set $\cpot=1$, $\varepsilon^2=0.00178$, and the horizon $\delta_{hf}=0.25$ of the high-fidelity model, which leads to $\xi = \cker-\cpot\approx 0.0253$, where $\cker\approx 36\varepsilon^2/\delta^2$.
We also set $\beta_1 = 1$ and $\beta_2= 0.05$ for the non-mass-conserving nonlocal Cahn-Hilliard model. 

\begin{figure}[h!]
\begin{center}
\begin{tabular}{ccc}
\includegraphics[height=1.5in]{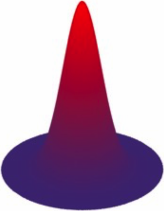}
&
\includegraphics[height=1.5in]{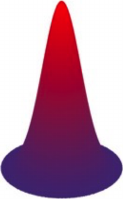}
&
\includegraphics[height=1.5in]{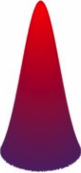}
\\
\small{$\delta=\delta_{hf}=0.25$}
&
\small{$\delta=0.75\delta_{hf}=0.1875$}
&
\small{$\delta=0.5\delta_{hf}=0.125$}
\\[-4ex]
\end{tabular}
\end{center}
\caption{The different values for the horizons govern the radius of the compact support for the kernel. Left to right: the kernel defined in \eqref{eq:kernel_gauss} have support over the discs having horizons $\delta=0.25,\, 0.1875,\, 0.125$, respectively. The peak value of the kernels for all three horizons is the same and is dictated by $\delta_{hf}$ as evident from \eqref{eq:kernel_gauss}. 
}
\label{fig:kernel}
\end{figure}

Furthermore, for the numerical experiments, we use three values for the horizon
$\delta=\delta_{hf}=0.25$, $\delta=0.75\delta_{hf}=0.1875$, and $\delta=0.5\delta_{hf}=0.125$ so that the supports of the kernel \eqref{eq:kernel_gauss} are the discs centered at $\x$ with horizons $0.25$, $0.185$, and $0.125$, respectively.
In Figure \ref{fig:kernel}, we illustrate the three kernels so defined. 

\begin{figure}[h!]
\begin{center}
\begin{tabular}{ccc}
\includegraphics[height=1.5in]{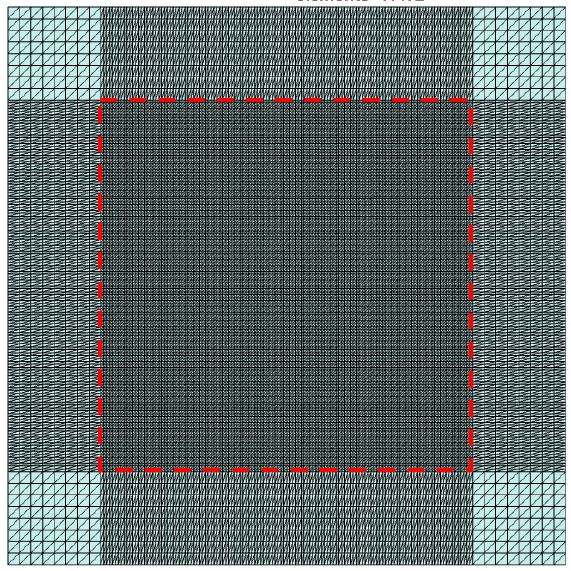}
&
\includegraphics[height=1.5in]{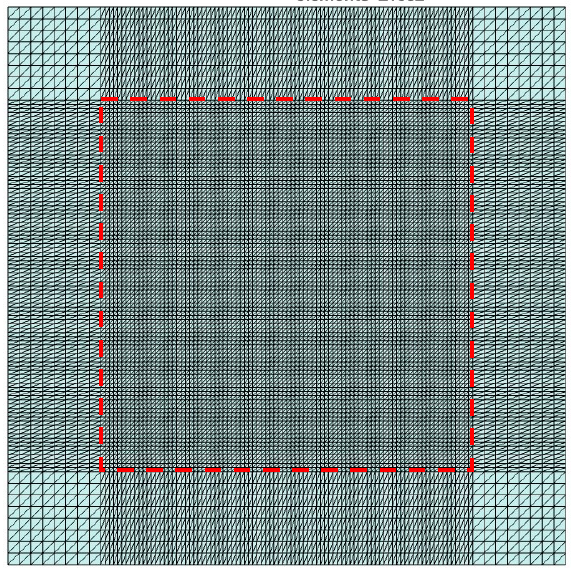}
&
\includegraphics[height=1.5in]{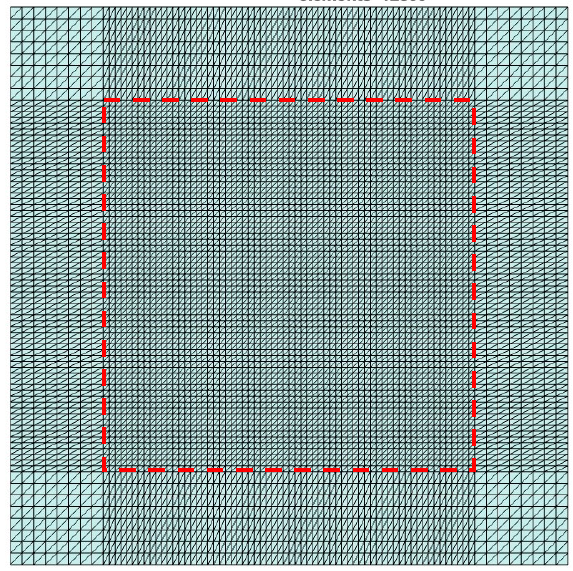}
\\
\small{$h=h_{hf}=1/128$}
&
\small{$h=1/88$}
&
\small{$h=1/64$}
\\[-4ex]
\end{tabular}
\end{center}
\caption{The figures represent the three levels of mesh refinement with the left figure corresponding to the high-fidelity mesh. The dashed red box represents the extent of the physical domain $\Omega$ and the surrounding domain represents the interaction domain $\Omega_I$. 
}
\label{fig:mesh}
\end{figure}

Nine models $\{f_i\}_{i=1}^9$ are defined using approximations of the solution $u(\x,T;\zz)$ for $T=1$ based on the pairing of three mesh sizes for the physical domain $h=h_{hf}=1/128$, $h=1/88$, and $h=1/64$ with three horizons $\delta=\delta_{hf}=0.25$, $\delta=0.1875$, and $\delta=0.125$. The mesh refinement in the interaction domain $\Omega_I$ is less refined as can be seen in Figure \ref{fig:mesh}.
We set a time step $dt=0.01$ for all models for the temporal discretization. According to \eqref{eq:kernel_gauss}, the magnitude of the horizon will only affect the volume of the compact support, yet the peak value of the kernel $\gamma(\x,\xx)$ is influenced by the magnitude of the high-fidelity horizon only. The pair $h=h_{hf}=1/128$ and $\delta=\delta_{hf}=0.25$ defines the high-fidelity model $f_{hf}=f_1$ and the other 8 pairs define the surrogate models. Note that coarser meshes and/or smaller horizons both result in the discretized nonlocal operator having increased sparsity and therefore lower solution costs for the system \eqref{eq:nonl_CH_nm} compared to that for the high-fidelity model.

\begin{figure}[h!]
\begin{center}
\begin{tabular}{ccc}
\includegraphics[width=.3\textwidth]{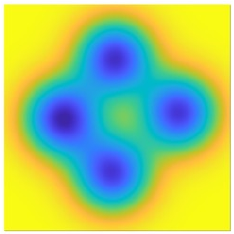}
&
\includegraphics[width=.3\textwidth]{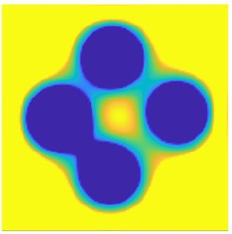}
&
\includegraphics[width=.3\textwidth]{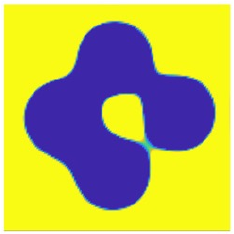}
\\
(a) $t=0$ & (b) $t=0.1$ & (c) $t=0.2$ \\
\includegraphics[width=.3\textwidth]{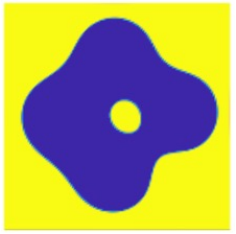}
&
\includegraphics[width=.3\textwidth]{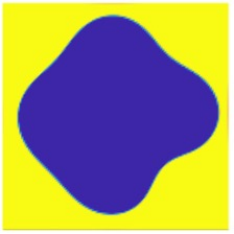}
&
\includegraphics[width=.3\textwidth]{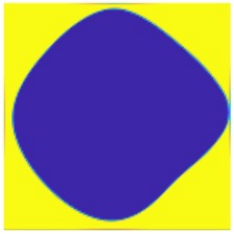}
\\
(d) $t=0.4$ & (e) $t=0.6$ & (f) $t=T=1.0$
\end{tabular}
\end{center}
\caption{(a): A sample initial condition with randomness in the location and amplitude of the four nuclei, where the colors blue and yellow correspond to the two distinct phases $u=-1$ and 1, respectively. (b) and (c): The evolution of the phases at five later times ($t=0.1,\, 0.2,\, 0.4,\, 0.6,\,$ and $t=T=1.0$, respectively). Note that the nonlocal generalized Cahn-Hilliard model is capable of capturing the evolution of the phases with sharp interfaces. 
}
\label{fig:time_history}
\end{figure}

The initial condition consists of four nuclei of phase $\p=-1$ determined from
\begin{equation}\label{eq:ic}
  u_0\left(\x \right) = 1-2\sum_{i=1}^4 \mu_i e^{-36|\x-\overline{\x}_i-{{\boldsymbol\eta}_i|^2}}, 
\end{equation}
where $\overline{\x}_1=\left[0.25,\, 0.5\right]^\top$, $\overline{\x}_2=\left[0.75,\, 0.5\right]^\top$, $\overline{\x}_3=\left[0.5,\, 0.25\right]^\top$, and $\overline{\x}_4=\left[0.5,\, 0.75\right]^\top$, and
the random parameter inputs are four independently, identically, and uniformly distributed realizations of $\zz_i=(\mu_i,{\boldsymbol\eta}_i)$, $i=1,2,3,4$, within the parameter input domain ${\boldsymbol\Theta}=\left[0.9\; 1.0\right] \times \left[-0.025\; 0.025\right]\times \left[-0.025\; 0.025\right] \subset \mathbb{R}^3$. The initial condition depicted in Figure \ref{fig:time_history}a is for the specific random samples $\mu_1=0.9815,{\boldsymbol\eta}_1=[0.0072,-0.0039]$, $\mu_2=0.9276,{\boldsymbol\eta}_2=[0.0232,0.0208]$, $\mu_3 = 0.9162,{\boldsymbol\eta}_3=[-0.0220,0.0041]$, and $\mu_4=0.9417,{\boldsymbol\eta}_4=[-0.0099,0.0118]$. The other two figures in Figure \ref{fig:time_history} depict the evolution of the interface at later times.

MFMC estimation requires knowledge of the Pearson correlation coefficients $\rho_{1,k}$, costs $\CC_k$, and variances $\sigma_k^2$ of the 9 models $\{f_{hf}=f_1,f_2,f_3,f_4,f_5,f_6,f_7,f_8,f_9\}$ which ensue from the 9 different pairs of $h$ and $\delta$; these pairs are given in the second and third columns of the same rows of Table \ref{tab:cost_rhoa}. We determine approximations of $\rho_{1,k}$, $\CC_k$, and $\sigma_k^2$ by using the average of fifty pilot samples. Note that the fourth column of Table \ref{tab:cost_rhoa} orders the models according to decreasing correlation coefficient $\rho_{1,k}$ as is required by \eqref{eq:conditiona}. 

\begin{table}[h!]
\begin{center}
\caption{Estimated Pearson correlation coefficients, costs, and standard deviations of the high-fidelity model and the 8 surrogate models $\{f_{hf}=f_1,f_2,f_3,f_4,f_5,f_6,f_7,f_8,f_9\}$ as determined from 50 random parameter samples $\zz$. The cost incurred by a single sampling of the high-fidelity model is $\CC_1=63.089$ seconds.}
\label{tab:cost_rhoa}
\begin{tabular}{ccccccc}
\toprule
 & mesh &  & correlation  & cost ratio & standard 
\\
models $f_k$ & width $h$ & horizon $\delta$ & coefficient $\rho_{1,k}$ &  $\CC_k/\CC_1$ & deviation $\sigma_k$
\\
\midrule
 $f_1=f_{hf}$ & $1/128$ & $0.25$ & 1.0000000 & 1.0000 & $6.694\times 10^{-3}$
\\
 $f_2$ & $1/128$ & $0.1875$ & 0.9999999 & 0.7501 & $6.679\times 10^{-3}$
\\ 
 $f_3$ & $1/88$ & $0.1875$ & 0.9999613 & 0.2514 & $6.604\times 10^{-3}$
\\
 $f_4$ & $1/88$ & $0.25$ & 0.9999612 & 0.3069 & $6.620\times 10^{-3}$
\\
 $f_5$ & $1/128$ & $0.125$ & 0.9999465 & 0.5214 & $6.494\times 10^{-3}$
\\
 $f_6$ & $1/88$ & $0.125$ & 0.9998767 & 0.2054 & $6.437\times 10^{-3}$
\\
 $f_7$ & $1/64$ & $0.1875$ & 0.9998020 & 0.1035 & $6.506\times 10^{-3}$
\\
 $f_8$ & $1/64$ & $0.25$ & 0.9998010 & 0.1204 & $6.525\times 10^{-3}$
\\
 $f_9$ & $1/64$ & $0.125$ & 0.9996606 & 0.0888 & $6.331\times 10^{-3}$
\\
\bottomrule
\vspace{-6ex}
\end{tabular}
\end{center}
\end{table}

Ensuing from the discussion in Section \ref{modelselection}, given the high-fidelity model and the eight surrogate models, we can define $2^8$-1=255 distinct subsets that include the high-fidelity model and anywhere from one to eight surrogate models. Among those 255 subsets, only 131 of them satisfy both conditions \eqref{eq:conditiona} and \eqref{eq:conditionb}. Table \ref{tab:p_condition} contains the first 10 and the last six subsets satisfying both conditions as well as two additional subsets, all of which are ordered from smallest to largest variance reduction ratio $\VVVt$ as is listed in the third column of that table. The values of the normalized minimum required budget $\ppp_{\text{min}}/C_1$ for each subset are listed in the fourth column of Table \ref{tab:p_condition}. The fifth column in each row gives the index of the subset occupying that row if the subsets were instead ordered in increasing minimum required budget. The first two highlighted rows depict the subsets having the lowest and second lowest variance reduction ratios $V$, respectively. The third highlighted row depicts the subset having the lowest required minimum budget $\ppp_{\text{min}}$. The fourth and fifth highlighted rows, respectively, depict the subsets consisting of the three grid values and the subset consisting of the three horizon values. A few observations can be deduced from Table \ref{tab:p_condition}:

\begin{table}[h!]
\begin{center}
\caption{Out of the $2^8-1=255$ possible distinct subsets, 131 of them are feasible, i.e., satisfying both conditions \eqref{eq:conditiona} and \eqref{eq:conditionb}. The table provides data for a few of the 131 retained subsets. The rows are ordered according to the increasing variance reduction ratio. The highlighted rows are those chosen to be discussed in the text and used in the numerical studies.
}
\label{tab:p_condition}
\begin{tabular}{clccc}
        \toprule
indexed   &         & variance  & normalized                & indexed \\
according to & \multicolumn{1}{c}{subsets\phantom{xxx}} & reduction & minimum required          & according \\
the value of $V$    &         & ratio $V$ & budget $\ppp_{{min}}/C_1$ & to $\ppp_{{min}}$ \\
        \midrule
        \rowcolor{light-gray}
        1 & $\SSS_1 =\{f_1,f_2,f_3,f_9\}$& 0.10122 & 794.55 & 72\\
        \rowcolor{light-gray}
        2 &  $\SSS_2 =\{f_1,f_3,f_9\}$ & 0.10172 & 36.23 & 22\\
        3 &  $\SSS_3 =\{f_1,f_2,f_4,f_9\}$ & 0.10205 & 797.79 & 73\\
        4 &  $\SSS_4 =\{f_1,f_2,f_3,f_8,f_9\}$ & 0.10254 & 799.70 & 74\\
        5 &  $\SSS_5 =\{f_1,f_4,f_9\}$ & 0.10255 & 36.33 & 23\\
        6 &  $\SSS_6 =\{f_1,f_2,f_3,f_7,f_9\}$ & 0.10281 & 800.79 & 75\\
        7 &  $\SSS_7 =\{f_1,f_2,f_9\}$ & 0.10291 & 801.17 & 76\\
        8 &  $\SSS_8 =\{f_1,f_2,f_8,f_9\}$ & 0.10297 & 801.37 & 77\\
        9 &  $\SSS_9 =\{f_1,f_3,f_8,f_9\}$ & 0.10304 & 36.47 & 24\\
        10 & $\SSS_{10} =\{f_1,f_2,f_4,f_8,f_9\}$ & 0.10314 & 802.07 & 78\\
        $\vdots$ & $\vdots$ & $\vdots$ & $\vdots$ & $\vdots$\\
        \rowcolor{light-gray}
        28 & $\SSS_{28} =\{f_1,f_9\}$ & 0.10491 & 12.43 & 1\\
        $\vdots$ & $\vdots$ & $\vdots$ & $\vdots$ & $\vdots$\\
        \rowcolor{light-gray}
        93 & $\SSS_{93} =\{f_1,f_4,f_8\}$ & 0.13372 & 41.49 & 55\\
        $\vdots$ & $\vdots$ & $\vdots$ & $\vdots$ & $\vdots$\\
        \rowcolor{light-gray}
        126 & $\SSS_{126} =\{f_1,f_2,f_5\}$ & 0.53494 & 1826.59 & 129\\
        127 & $\SSS_{127} =\{f_1,f_5\}$ & 0.53639 & 70.82 & 69\\
        128 & $\SSS_{128} =\{f_1,f_2,f_3,f_5\}$ & 0.53697 & 1830.06 & 130\\
        129 & $\SSS_{129} =\{f_1,f_3,f_5\}$ &  0.53813 & 83.34 & 71\\
        130 & $\SSS_{130} =\{f_1,f_4,f_5\}$ & 0.53855 & 83.26 & 70\\
        131 & $\SSS_{131} =\{f_1,f_2\}$ & 0.75076 & 2613.90 & 131\\
        \bottomrule
        \vspace{-6ex}
\end{tabular}
\end{center}
\end{table} 

\begin{itemize}[topsep=0pt,parsep=-1pt,leftmargin=*]
    \item[--] If the only criterion used for selecting a subset for the MFMC estimation is to minimize the variance reduction ratio $\VVV$, we would choose the subset $\{f_1,f_2,f_3,f_9\}$ corresponding to the first row in Table \ref{tab:p_condition}. However, the minimum required computational budget would have the rather large value $\ppp_{{\min}} = 794.55C_1 \approx$ 50,127 seconds.
    \item[--] On the other hand, the subset $\{f_1,f_3,f_9\}$ corresponding to the second row in Table \ref{tab:p_condition} results in a negligible $0.50\%$ increase in the variance reduction ratio $\VVV$, yet it achieves more than one order of magnitude reduction in the minimum required budget, i.e., we now have that $\ppp_{\text{min}}=36.23C_1 \approx$ 2,286 seconds. For the subset $\{f_1,f_2,f_3,f_9\}$, the correlation coefficient of model 2 with the high-fidelity model is very close to unity, thus reducing the denominator in \eqref{eq:bmin} significantly, which then results in a high value of $\ppp_{\text{min}}$ for this subset. On the other hand, for the subset $\{f_1,f_3,f_9\}$, the correlation coefficient of the model with the highest correlation with the high-fidelity model (model 3 in this case) is lower than that for model 2 and this reduces the minimum required budget.
    \item[--] Within the 131 subsets satisfying both conditions, subset $\{f_1,f_2\}$ consisting of the high-fidelity model and the highest-correlated surrogate model gives the least variance reduction. This is due to the fact that their relative cost ratio is significant, i.e., we have that $C_2/C_1=0.7501$.
\end{itemize}

Next, we provide numerical results for multifidelity estimators corresponding to the five highlighted subsets in Table \ref{tab:p_condition}.
\begin{itemize}[topsep=0pt,parsep=-1pt,leftmargin=*]
    \item[--] Case 1: {\em optimal variance reduction ratio}: $\SSS_1=\{f_1,f_2,f_3,f_9\}$ that has the lowest variance reduction ratio $\VVV_1=0.10122$ and has a rather large minimum required budget $\ppp_{\text{min}}^{(1)} = 794.55 C_1 =$ 50,127.4 seconds.
    \item[--] Case 2: {\em nearly optimal variance reduction ratio}: $\SSS_2=\{f_1,f_3,f_9\}$ that has the second lowest variance reduction ratio $\VVV_2=0.10172$ and has a much smaller minimum required budget $\ppp_{\text{min}}^{(2)} = 36.23 C_1 =$ 2,285.7 seconds compared to that of Case 1.
    \item[--] Case 3: {\em optimal budget}: $\SSS_{28}=\{1,9\}$ that has the smallest minimum required budget $\ppp_{\text{min}}^{(28)} = 12.43 C_1 = 784.2$ seconds and has the variance reduction ratio $\VVV_{28}=0.10491$ that is within $3.65\%$ of the best variance reduction ratio of Case 1.
    \item[--] Case 4 -- {\em three-$h$}: $\SSS_{93}=\{1,4,8\}$ having the three models with different grid sizes and the fixed high-fidelity horizon $\delta_{hf}$ for which $\VVV_{93}=0.13372$ and $\ppp_{{min}}^{(93)} = 41.49 C_1 =$ 2,617.6 seconds.
    \item[--] Case 5 -- {\em three-$\delta$}: $\SSS_{126}=\{1,2,5\}$ having the three models with different horizons and the fixed high-fidelity grid size $h_{hf}$ for which $\VVV_{126}=0.53494$ and $\ppp_{\text{min}}^{(126)} =1826.59  C_1 =$ 115,237.7 seconds.
\end{itemize}
For comparison purposes, we also consider the case given by:
\begin{itemize}[topsep=0pt,parsep=-1pt,leftmargin=*]
    \item[--] Case MC -- {\em Monte Carlo estimation}: the estimator \eqref{eq:mc} for the single model having grid size $h_{hf}$ and horizon $\delta_{hf}$.
\end{itemize}

\vskip5pt
\noindent
{\bf Remark.} We note that cases 4, and 5 fall within the purview of the popular {\em multilevel} Monte Carlo (MLMC) methods \cite{Heinrich_2001,Giles_2015} by which we mean that the high-fidelity model and the surrogate models differ only in the values of a single parameter, e.g., for Case 4, that parameter is the grid size.
\quad$\Box$

To reduce the minimum required computational budget for subset $\SSS_k$, i.e., $\ppp_{\text{min}}^{(k)}$, even further, the remedy introduced in \cite{Gruber_Gunzburger_Ju_Wang_2022} is utilized: for budgets $B<\ppp_{\text{min}}^{(k)}$ (resulting in $m_1<1$), the following steps are taken to determine the number of samples for subset $k$ with $|\SSS_k|$ models (Algorithm 2 in \cite{Gruber_Gunzburger_Ju_Wang_2022} with slight modification):
\begin{enumerate}[leftmargin=*,label=\arabic*:,parsep=0pt]
    \item Determine the number of samples according to \eqref{eq:samples}.
    \item {\bf while} there is $1\leq j \leq|\SSS_k|-1$ with $m_j<j$ {\bf do}
    \item \qquad Select the smallest index $j$ with $m_j<j$. Set $m_j=j$ and $r_{j+1}=1$.
    \item \qquad For $j+1 \leq \ell \leq |\SSS_k|$, determine the oversampling ratio $r_\ell$ 
    \begin{equation}
        r_\ell = \left(\dfrac{C_{i_{j+1}}(\rho_{1,i_\ell}^2-\rho_{1,i_{\ell+1}}^2)}{ C_{i_{\ell}}(\rho_{1,i_{j+1}}^2-\rho_{1,i_{j+2}}^2)}\right)^{1/2},
    \end{equation}
    \item \qquad Determine the number of samples $m_{\ell}$ for $j+1 \leq \ell \leq |\SSS_k|$,
    \begin{equation}\label{eq:below_min_budget}
        m_{j+1}=\frac{B-\sum_{\ell=1}^{j}\ell C_{i_\ell}}{\sum_{\ell=j+1}^{|\SSS_k|} {C_{i_\ell} r_\ell}}, \quad m_\ell=m_{j+1}r_{\ell},\;\; \text{for}\; j+1<\ell\leq |\SSS_k|
    \end{equation}
\end{enumerate}
In this study, we enforce the requirement $B\geq\sum_{\ell=1}^{|\SSS_k|} \ell C_{i_{\ell}}$ for budgets below the minimum required budget $\ppp_{\text{min}}^{(k)}$ for subset $\SSS_k$, to ensure non-zero contributions in the telescopic sum of \eqref{eq:MFMC_estimator}. This is a stricter requirement than that of \cite{Gruber_Gunzburger_Ju_Wang_2022}, i.e., $B>\sum_{\ell=1}^{|\SSS_k|} C_{i_{\ell}}$.

We now report on the numerical results related to MFMC estimation for the subsets listed in Cases 1 through 5 with the views of comparing among those five cases and comparing all those cases to MC estimation. For this purpose, we choose 8 computational budgets between and including $B= 10^3$ and $10^{5\frac{1}{3}}$ which are uniformly sampled in a logarithmic scale. 
The lowest budget for each multifidelity case is determined according to \eqref{eq:bmin} and \eqref{eq:below_min_budget}.

\begin{figure}[h!]
\centerline{
\includegraphics[angle=-90,width=3.1in]{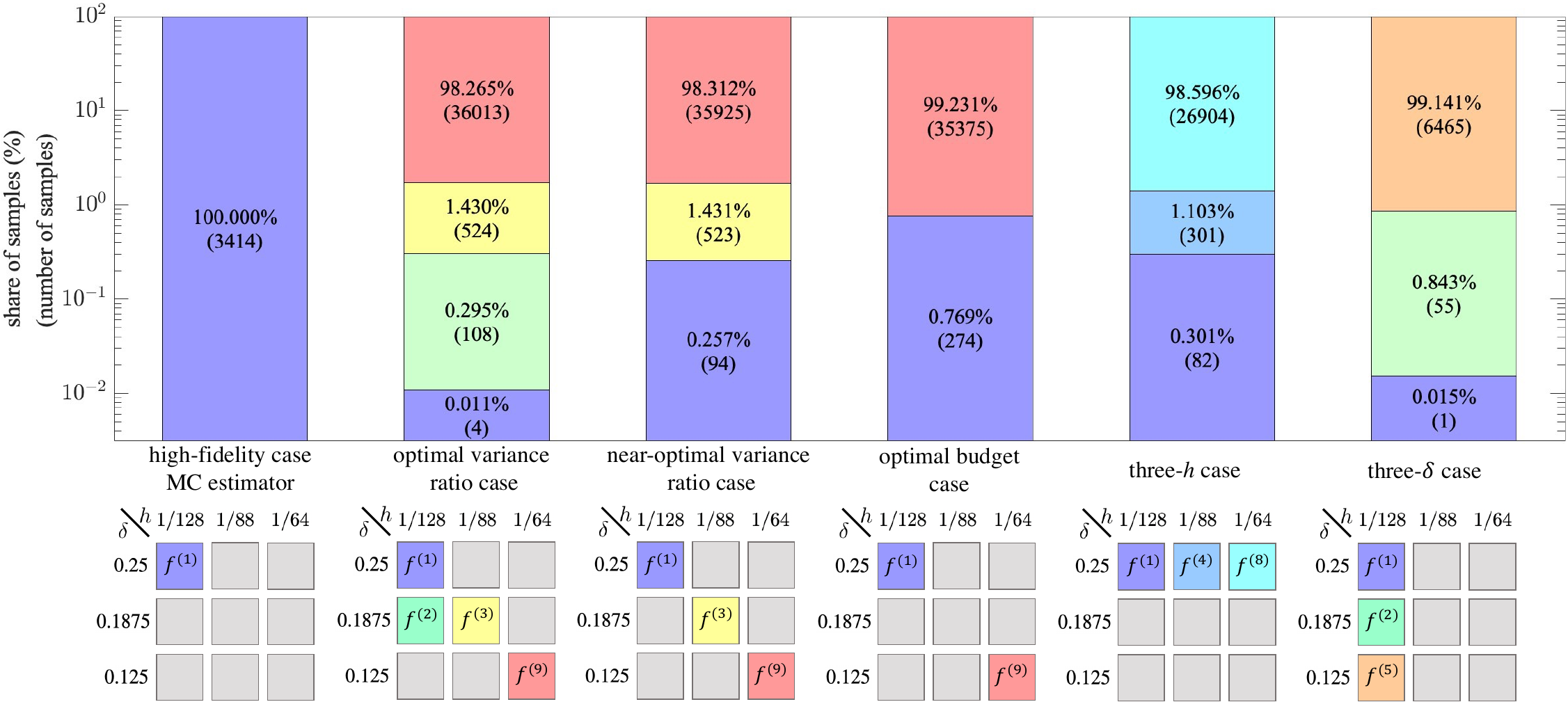}
}
\caption{The figure provides the percentage and actual number of samples (in parentheses) for each model $f_{i_j}$ within the five MFMC cases compared to that for the MC case that only queries the high-fidelity model.}
\label{fig:samples}
\end{figure}

The bottom row in Figure \ref{fig:samples} shows, from left to right, the members of the subsets of $\{f_1,\ldots,f_9\}$ contributing to Case MC and Cases 1 to 5. The columns in the first-row show, again from left to right for the MC and the five MFMC estimators, the percentage $m_j/\sum_{j=1}^{|\SSS|} m_j \times 100$ share and the actual number (in parentheses) of samples taken for each model $f_{i_j}$ within those cases. Note the logarithmic scale in the ordinate. The provided number of samples in Figure \ref{fig:samples} correspond to the {\it largest} budget $B = 10^{5\frac{1}{3}} \approx$ 215,443 seconds. The reasoning behind choosing this specific budget is to ensure it is larger than the minimum required budget $\ppp_{\text{min}}^{(k)}$ for all five multifidelity cases. For the multifidelity cases, most of the budget is assigned to the lower-cost surrogate models for speedup whereas the high-fidelity samples preserve unbiasedness. 

The MC and MFMC estimators are determined for several computational budgets and the MSE is used as the validation metric (as discussed in Section \ref{estimator32}). To determine an approximation $Q_{validation}$ of the true QoI $\QQ = {\mathbb E}[f_1(\zz)]$, we determine an MC estimator $\QQ_{validation}$ determined from 360,000 samples, i.e., we average the OoI $f_1(\zz)=f_{hf}(\zz)$ over that many samples to obtain
$$
\QQ_{validation} = \frac1{360000} \sum_{m=1}^{360000} f_1(\zz_m)
\approx
\QQ = \mathbb{E}\left[f_1(\mathbf{z})\right] .
$$
The approximate MSE \eqref{eq:mse_error} is determined for several\ computational budgets $B$ and are plotted in Figure \ref{fig:MSE}. In that figure, the dashed lines depict the theoretical decay of the MSE with increasing budget as determined from \eqref{eq:error_theory} for each case. The solid lines depict the approximate MSEs as determined from the numerical simulations 
including empty and filled circle markers. The empty circle markers correspond to data points with budgets above the minimum required budget $B>B_{\min}^{(k)}$ whereas the filled circle markers represent the data points with budgets below $B_{\min}^{(k)}$ for which the number of samples is determined according to \eqref{eq:below_min_budget}. The following observations can be made from Figure \ref{fig:MSE}:
\begin{figure}[h]
    \centering
    \includegraphics[width=0.7\textwidth]{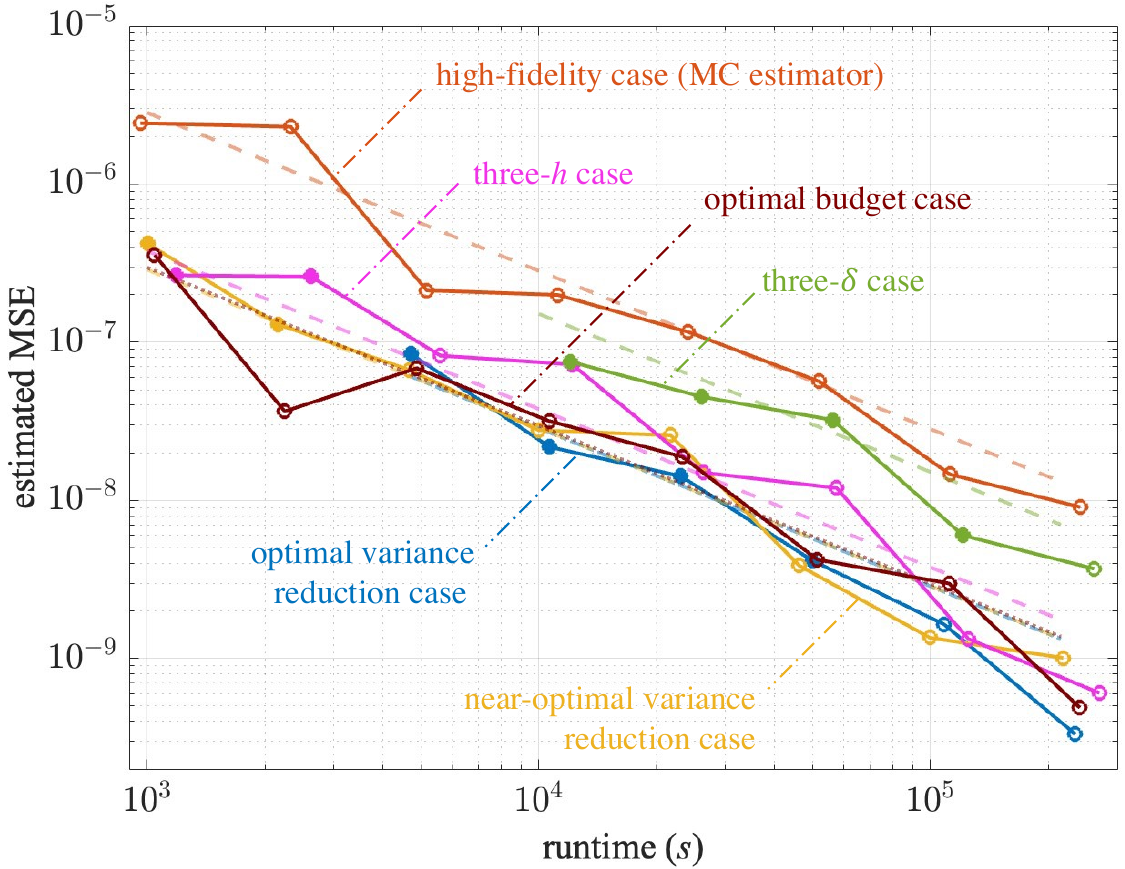}
   \caption{The solid lines in the figure display the decay of the estimated MSE with increasing runtime for the different cases. The dashed lines correspond to the theoretical ${\rm MSE}_{theoretical}$ as determined from \eqref{eq:error_theory}. The required minimum budget for each case is determined according to the discussions in Section \ref{section4}. 
   The filled circle markers in the plots for each case correspond to computational budgets below the minimum required budget for that case, and the corresponding number of samples is determined according to \eqref{eq:below_min_budget}. The empty circle markers are for budgets above the minimum required budget.}
   \label{fig:MSE}
\end{figure}

\begin{itemize}[topsep=0pt,parsep=-2pt,leftmargin=*]
    \item[--] For the same value of the budget, all five MFMC estimators achieve lower values of the MSE relative to that of the MC estimator.
    \item[--] For the same value of the MSE, all five MFMC estimators require lower budgets relative to that of the MC estimator.
    \item[--] Similar to the observation made in \cite{Khodabakhshi_Willcox_Gunzburger_2021} for the nonlocal diffusion model, compared to the three-$\delta$ case (i.e., Case 5), the three-$h$ case (i.e., Case 4) shows significantly better performance at MSE reduction for a fixed budget and also for budget reduction for a fixed MSE.
    \item[--] The optimal variance reduction ratio, the near-optimal variance reduction ratio, and the optimal budget cases (i.e., Cases 1, 2, and 3, respectively) achieve, for a given computational budget, about an order of magnitude reduction in the MSE compared to the MC case. This remains to be true even for the lower values of the computational budget. These three cases, which include surrogate models having smaller horizons and larger grid sizes, are viable candidates for use as an MFMC estimator.
    \item[--] The marginal increase in the variance reduction ratio for the near-optimal variance reduction ratio case (Case 2) and optimal budget case (Case 3) compared to that of the optimal variance reduction ratio case (Case 1) are negligible in both the theoretical sense (i.e., we have coincident dashed lines) and in the numerical sense (i.e., we have the proximity of the solid lines). This observation would possibly lead one (should multiple surrogate models exist and the number of feasible subsets is greater than one) to also consider a few additional subsets having slight increases in their variance reduction ratios so as to examine their minimum required budget and the relative increase in the variance reduction ratio.
\end{itemize}

\section{Concluding remarks}\label{section5}

In this paper, we utilize the MFMC method to reduce the computational cost of estimating the expectation of an OoI depending on the solution of the generalized nonlocal Cahn-Hilliard model which itself depends on random parameters. The surrogate models used in the MFMC method are defined by using coarser grids and smaller horizons than those used to define the high-fidelity OoI. Using the MFMC method, for a given computational budget, we achieve about an order of magnitude reduction in the MSE of the MFMC estimator compared to that of the MC estimator. 
In problems with multiple surrogate models, several MFMC estimators can be constructed by taking distinct unique subsets of the models. Commonly, the subset with the highest variance reduction is chosen as the optimal subset for the MFMC estimator. 
For the application of interest in this study, it is shown that it is better to consider both the highest variance reduction and the minimum required budget for an unbiased estimator.
Specifically, simultaneous consideration of both conditions results in more than one order of magnitude reduction in the required computational budget while only slightly increasing the variance reduction ratio.
Similar to the nonlocal diffusion model \cite{Khodabakhshi_Willcox_Gunzburger_2021}, for the generalized nonlocal Cahn-Hilliard model, the MFMC estimator using surrogate models with larger grid sizes demonstrates better performance compared to the estimator using surrogate models with smaller horizons. However, the optimal subsets (optimal variance reduction, near-optimal variance reduction, and optimal budget cases) include surrogate models having both smaller horizons and larger grid sizes.

\vskip6pt

\noindent{\bf Acknowledgments}

This work was supported by US Department of Energy grants DE-SC0019303 and DE-SC0021077 as part of the AEOLUS Multifaceted Mathematics Integrated Capability Center at the University of Texas at Austin. 
The first author also acknowledges the availability of the San Diego Supercomputing Center funded by the NSF ACCESS. 
The second author acknowledges that the work was supported by the U.S. Department of Energy, Office of Advanced Scientific Computing Research, Applied Mathematics Program under the award number ERKJ345; and was performed at the Oak Ridge National Laboratory, which is managed by UT-Battelle, LLC under Contract No. De-AC05-00OR22725. 
Figure \ref{fig:figure1} that appears in \cite{Burkovska_Gunzburger_2021} is used with permission from the publisher of that paper.

\vspace{-6pt}
\bibliographystyle{elsarticle-num-names} 
\bibliography{refs2}

\end{document}